\documentclass[10pt,a4paper]{article}
\usepackage{amsmath}
\usepackage{amsfonts}
\usepackage{amssymb}
\usepackage{graphicx}
\usepackage{textcomp}
\usepackage{url}
\usepackage{lipsum}
\usepackage{setspace}
\usepackage{comment}
\usepackage{color}
\usepackage{pdfpages}
\usepackage[numbers,sort&compress]{natbib}
\usepackage[usenames,dvipsnames,svgnames,table]{xcolor}
\usepackage[lofdepth,lotdepth,caption=false]{subfig}
\usepackage[pdfstartview=FitH]{hyperref}
\usepackage{hyperref}
\newsavebox{\bigleftbox}
\hypersetup{
 colorlinks,
 citecolor=Blue,
 linkcolor=Blue,
 urlcolor=Blue
}

\makeatletter
 \def\footnoterule{\kern-3\p@
   \noindent\hrulefill \kern 2.8\p@} 
\makeatother
\usepackage[left=3.00cm, right=3.00cm, top=2.00cm, bottom=2.00cm]{geometry}

\usepackage{fancyhdr}

\title{\textbf{TODD-Graphene: A Novel Porous 2D Carbon Allotrope for High-Performance Lithium-Ion Batteries}}
\author{
    E. J. A. Santos$^{1,2,\P}$
    K. A. L. Lima$^{1,2,\P}$, 
	L. A. Ribeiro Junior$^{1,2,\S}$
	}
\date{}

\begin{document}
    \maketitle
	\vspace{-0.6cm}
	\begin{center}\small
	\textit{Institute of Physics, University of Bras\'ilia, 70910-900, Bras\'ilia, Brazil}\\
            \textit{Computational Materials Laboratory, LCCMat, Institute of Physics, University of Bras\'ilia, 70910-900, Bras\'ilia, Brazil}\\
		\phantom{.}\\ \hfill
        $^{\P}$\url{emanueljose1215@hotmail.com}\hfill
        $^{\P}$\url{kleutonantunes@gmail.com}\hfill
		$^{\S}$\url{ribeirojr@unb.br}\hfill
		\phantom{.}
	\end{center}
	

\onehalfspace

\noindent\textbf{Abstract: The class of 2D carbon allotropes has garnered significant attention due to its exceptional optoelectronic and mechanical properties, crucial for diverse device applications, such as energy storage. This study employs density functional theory calculations, ab initio molecular dynamics (AIMD), and classical reactive (ReaxFF) molecular dynamics (MD) simulations to introduce TODD-Graphene, a novel 2D planar carbon allotrope with a porous structure composed of 3-8-10-12 carbon rings. TODD-G exhibits intrinsic metallic properties with low formation energy and demonstrates exceptional dynamic, thermal, and mechanical stability. Calculations reveal a high theoretical capacity for adsorbing Li atoms by showing a low average diffusion barrier of 0.83 eV and a metallic framework boasting excellent conductivity, emerging as a promising anode material for lithium-ion batteries. We also calculated the charge carrier mobility for electrons and holes in TOOD-G, and the values surpassed the graphene ones. Classical reactive MD simulation results suggested its structural integrity with no bond reconstructions at 1800 K.}

\section{Introduction}

Since the discovery of graphene in 2004 \cite{Novoselov, Geim}, several optical, mechanical, electronic, and thermodynamic studies have been conducted on this material, revealing its unique underlying properties, most related to its atom-thick 2D honeycomb-like arrangement. From that year on, various new 2D carbon-based allotropes have been computationally proposed \cite{enyashin2011graphene,wang2015phagraphene,lu2013two,zhang2015penta,terrones2000new,PhysRevB.70.085417,wang2018popgraphene,tromer2023mechanical,junior2023irida}. Some of them have been recently synthesized, like the 2D biphenylene network \cite{Qitang}, the multilayer $\gamma-$graphyne \cite{Desyatkin}, and the monolayer fullerene network \cite{hou2022synthesis}. Nevertheless, due to the successful synthesis of these carbon allotropes, the search for new materials capable of revolutionizing flat electronics has intensified even further.

One of the current trends in research on new 2D carbon allotropes is the quest for structures that differ structurally from graphene, such as those with large pores \cite{zheng2015two,borchardt2017toward,tao2020advanced,fang20222d,zhang2013controlling,macha20192d,yao2018scalable,kochaev2023ionic,dolina2023thermal,mortazavi2023electronic,mortazavi2023theoretical}. This trend is because nanomaterials with distinct and broader ring structures demonstrate a superior capacity for lithium atom absorption compared to graphene \cite{Yu,kim2018study,li,wang2019planar,Da,zhang2023li,mortazavi2022electronic}. Examples of these structures with atomic thickness also include the pop-graphene \cite{wang2018popgraphene} with five, eight, and five-atom rings, 8-16-4-Graphyne \cite{tromer2023mechanical}, Irida-graphene \cite{junior2023irida} with five, eight, and six-atom rings, and 2D biphenylene network \cite{Qitang} with periodically arranged four-, six-, and eight-membered rings of sp$^2$-hybridized carbon atoms. Drawing from the results regarding Li storage capacity, proposing additional porous 2D carbon-based materials featuring interconnected porous rings could offer new channels for energy storage applications.

In 2D materials, the band gap is often highly responsive to external stimuli such as tensile stress or compressive strain \cite{peng2020strain}. This tunability has been a critical feature in tailoring the electronic properties of various 2D materials for diverse applications \cite{heine2015transition,chaves2020bandgap,zeng2015band,kulyamin2023electronic}. In this context, the quest for unconventional 2D materials that can resist significant changes in their band structure under moderated stress regimes is actual. Such materials, capable of maintaining their optoelectronic properties despite external stress, hold significant promise for developing robust and versatile components in flexible optoelectronics. Consequently, this study aims to contribute to exploring this possibility.

In this work, we used a computational protocol to propose a new 2D carbon allotrope formed by 3-8-10-12-membered rings of sp$^2$-hybridized carbon atoms, referred to as TODD-Graphene (TODD-G, see Figure \ref{fig:system}) from a bottom-up approach. We conducted density functional theory (DFT) and ab initio molecular dynamics (AIMD) simulations to study its electronic, optical, and mechanical properties. Our findings show that TODD-G is metallic and structurally stable, as indicated by its integrity at 1800K. The absence of imaginary phonon modes in its phonon dispersion profile also suggests its dynamical stability. This material presents optical activity in the visible and ultra-violet regions. TOOD-G has a low average diffusion barrier (about 0.85 eV) and a metallic framework boasting excellent conductivity, emerging as a promising anode material for lithium-ion batteries. Moreover, classical reactive MD simulation results suggested its structural integrity with no bond reconstructions at 1800 K.

\section{Metodology}

The CASTEP code \cite{ClarkSegal}, as implemented in Biovia Materials Studio software \cite{systemes2017biovia}, was used to perform the DFT and AIMD simulations, focusing on investigating TODD-G's mechanical, electronic, thermodynamic, and optical properties. Exchange-correlation functionals were treated using the generalized gradient approximation (GGA). Specifically, we employed both the Perdew–Burke –Ernzerhof (PBE) \cite{perdew1996generalized} and the hybrid Heyd–Scuseria–Ernzerhof (HSE06) \cite{heyd2003hybrid} functionals. To account for the interactions between the nuclear electrons, we used norm-conserving pseudopotentials implemented in CASTEP.

To achieve electronic self-consistency, we implemented the nonlinear iterative Broyden-Fletcher-Goldfarb-Shanno (BFGS) algorithm \cite{HEAD1985264, PFROMMER1997233}. We used a plane-wave basis set with an energy cutoff of 600 eV and a convergence criteria for energy of 1.0 × 10$^{-5}$ eV. When relaxing the TODD-G lattice, we applied periodic boundary conditions, ensuring that the residual force on each atom was less than 1.0 × 10$^{-3}$ eV/\r{A}, and the pressure remained below 0.01 GPa. 

The TOOD-G lattice was optimized by keeping the base vector fixed along the z-direction and employing a $10\times10\times1$ k-point grid. For electronic and optical calculations, we used $15\times15\times1$ k-point grids for the GGA/PBE method and $5\times5\times1$ for the HSE06 method, respectively. The partial density of states (PDOS) was calculated at the HSE06 level using a $20\times20\times1$ k-point grid. Elastic properties were determined using the local density approximation (LDA) within the scope of the LDA/CA-PZ method \cite{PhysRevLett.45.566,PhysRevB.23.5048}. We also considered a 15 \r{A} vacuum region to avoid spurious interactions between periodic images.

We employed a linear response method to analyze the phonon characteristics with a grid spacing of 0.05 \r{A} and a convergence tolerance of 10$^{-5}$ eV/\r{A}$^2$. To assess the mechanical properties of TODD-G, we adopted a stress-strain approach based on the Voigt-Reuss-Hill method \cite{zuo1992elastic, chung1967voigt}. Its stability was also tested through AIMD simulations. We employed an NVT ensemble with a fixed time step of 1.0 fs for a total duration of 5 ps. The Nosé-Hoover thermostat \cite{nose1984unified} was used to control the system’s temperature. In these simulations, a $2\times2\times1$ supercell comprising 56 atoms was used. These parameters were also used in other AIMD studies \cite{sangiovanni2018ab,lundgren2022perspective}.

An external electric field of 1.0 V/\r{A} was applied along the x, y, and z directions to investigate the optical properties of TODD-G. Using the complex dielectric constant $\epsilon=\epsilon_1+i\epsilon_2$, where $\epsilon_1$ and $\epsilon_2$ represent the real and imaginary parts of the dielectric constant, respectively, we directly obtained the optical quantities as described in reference \cite{lima2023dft}.

To determine the melting point of TODD-G, we employed fully atomistic reactive (ReaxFF) MD simulations. These simulations were conducted using the GULP module, implemented in Biovia Materials Studio software \cite{systemes2017biovia}, and considered the ReaxFF potential \cite{van2001reaxff}. This potential is widely recognized for its effectiveness in modeling nanostructure's mechanical properties and thermal stability due to its capacity to capture atomic-level bond formation and breakage \cite{mueller2010development}. 

The thermal stability of TODD-G was investigated by heating the system from 300 K up to 5000 K (heating ramp simulations). In the initial stages of the simulations, we minimized the energy of TODD-G. Subsequently, we subjected it to a thermostat chain to achieve thermodynamic equilibrium. Constant NPT ensemble integration at zero pressure and 300 K was performed to eliminate any remaining stress, employing a Nose/Hoover \cite{evans1985nose} pressure barostat for 50 ps. Following this, we coupled all systems in the canonical NVT ensemble for an additional 100 ps, allowing the generation of sampled positions and velocities at room temperature. The time step for these simulations was set to 0.05 fs.  

\section{Results}

\subsection{Structure and Stability}

\begin{figure}[!htb]
	\centering
	\includegraphics[width=\linewidth]{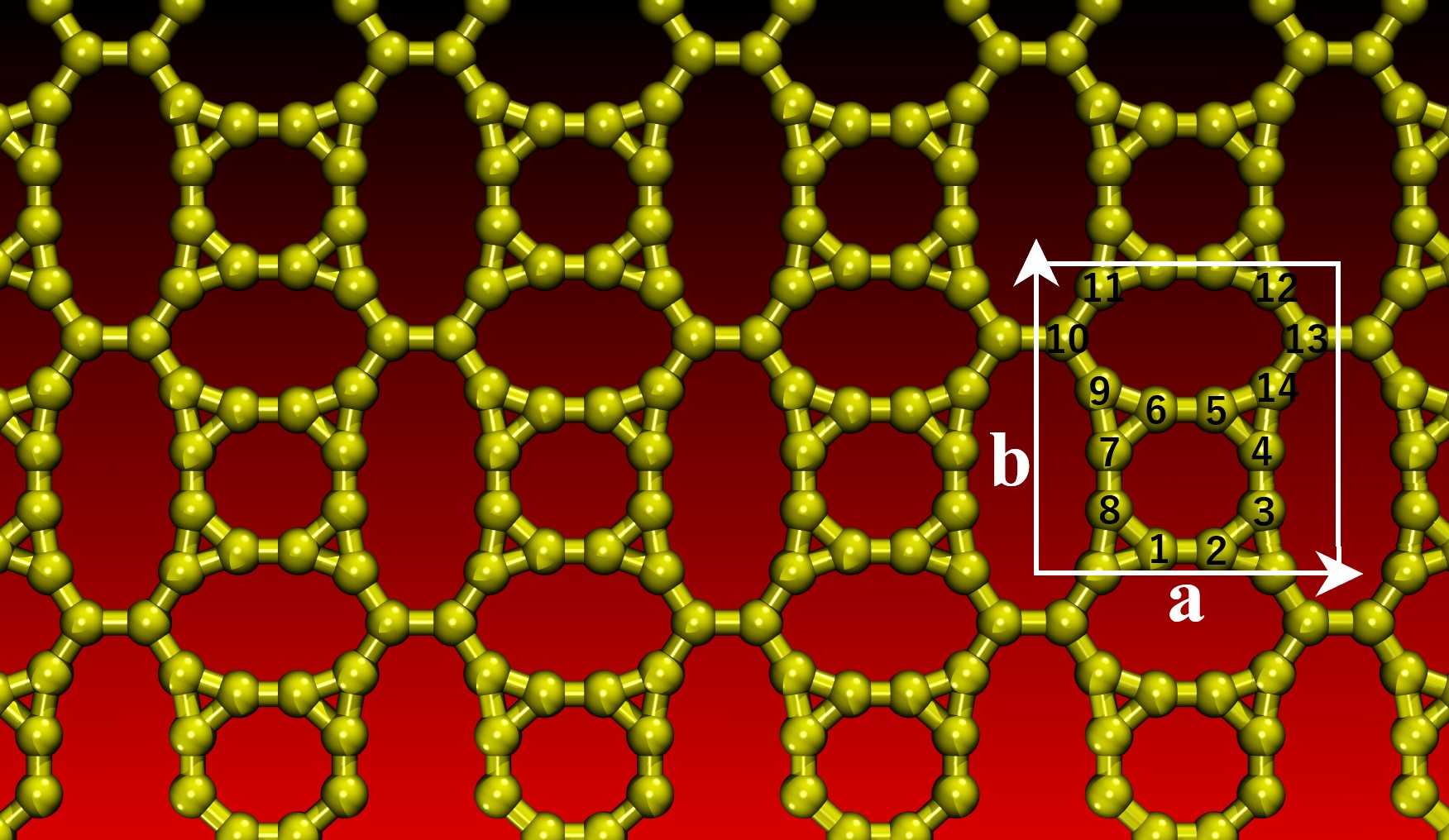}
	\caption{Schematic representation of the lattice topology for TODD-G. The white rectangle defined by lattice vectors a and b highlights its unit cell.}
	\label{fig:system}
\end{figure}

We begin our analysis by presenting the structural characteristics of TODD-G. The geometry optimization calculations produced consistent lattice parameters using the PBE and HSE06 methods. In this way, we emphasize only the results obtained within the HSE06 scheme. Figure \ref{fig:system} illustrates TODD-G's atomic arrangement and unit cell with related lattice vectors. The unit cell has 14 atoms, with dimensions $a=7.03$ \r{A} and $b=6.54$ \r{A}, and it exhibits an orthorhombic structure within the $P1$ space group. TODD-G forms a flat and periodic arrangement of sp$^2$-hybridized carbon atoms featuring interconnected 3-8-10-12-membered rings. See Table \ref{tab:01} for a more comprehensive view of its bond configuration. Despite the unique topology of TODD-G, it is worth noting that the interatomic distances in this material align with those found in other 2D carbon allotropes \cite{enyashin,Heimann1997CarbonAA}, underscoring its structural consistency.

It is worth noting that the planar density of TODD-G is 0.30 atom/\r{A}$^2$, surpassing that of graphyne (0.29 atom/\r{A}$^2$) \cite{peng2012mechanical} and graphdiyne (0.23 atom/\r{A}$^2$) \cite{valencia2006lithium}. However, it is lower than that of graphene (0.38 atom/\r{A}) \cite{PhysRevLett} and DHQ-graphene (0.33 atom/\r{A}$^2$) \cite{wang2019dhq}. The formation energy of TODD-G is -8.25 eV/atom, higher than that of graphene (-9.220 eV/atom) \cite{PhysRevB}, implying that the former is less stable than the latter.

\begin{table}[!htb]
    \centering
    \caption{Bond distances for the atoms highlighted in Figure \ref{fig:system}.}
    \label{tab:01}
    \begin{tabular}{|c|c|c|c|c|}
        \hline
        \textbf{Bond Type} & \textbf{Bond Length (\r{A})} & \textbf{Bond Type} & \textbf{Bond Length (\r{A})}\\
        \hline
        C1--C2  & 1.368 & C9--C7   & 1.435  \\ 
        \hline
        C2--C3  & 1.424 & C9--C6   & 1.392  \\
        \hline            
        C3--C4  & 1.368 & C9--C10  & 1.410  \\
        \hline            
        C4--C5  & 1.424 & C10--C11 & 1.410  \\
        \hline            
        C5--C6  & 1.368 & C12--C13 & 1.410  \\
        \hline            
        C6--C7  & 1.424 & C13--C14 & 1.410  \\
        \hline            
        C7--C8  & 1.368 & C5--C14  & 1.392  \\
        \hline            
        C1--C8  & 1.424 & C4--C14  & 1.435  \\   
        \hline
    \end{tabular}
\end{table}

\begin{figure}[!htb]
	\centering
	\includegraphics[width=\linewidth]{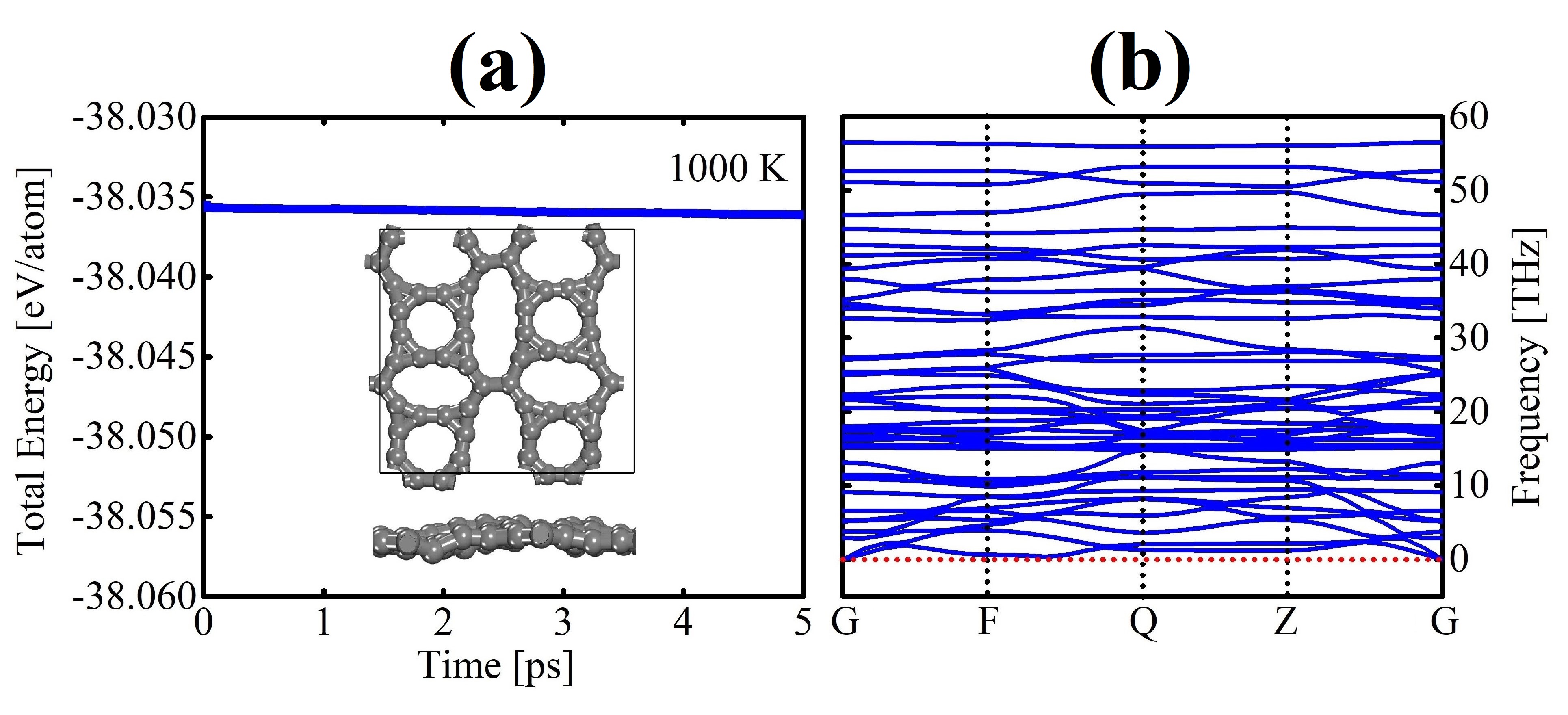}
	\caption{(a) Time evolution of the total energy per atom lattice at 1000K and (b) phonon band structure of the TOOD-G, both calculated at PBE level. The insets in panel (a) show the top and side views of the final AIMD snapshot at 5 ps.}
	\label{fig:stability}
\end{figure}

To evaluate the thermal stability of TODD-G, we conducted AIMD simulations as illustrated in Figure \ref{fig:stability}(a). In these simulations, we monitored the temporal variations in the total energy per atom during 5 ps, subjecting the material to a temperature of 1000 K. One can observe that variations in the total energy present a nearly flat pattern with minimal fluctuations. The MD snapshots within Figure \ref{fig:stability}(a) reveal slight deviations in TODD-G's initial planarity and bond distances caused by the elevated temperature without bond breaks or reconstructions. The final topology of TODD-G at 1000 K remains consistent with the optimized structure (refer to Figure \ref{fig:system}).   

To further assess the dynamical stability of TODD-G, we performed phonon dispersion calculations, as depicted in Figure \ref{fig:stability}(b). In this figure, one can not observe imaginary frequencies, indicating the inherent dynamical stability of TODD-G. The absence of a band gap between acoustic and optical modes suggests a significant scattering rate and relatively shorter phonon lifetimes, contributing to a moderate lattice thermal conductivity in this material. A well-established principle is that higher phonon frequencies correspond to stronger chemical bonds. In the case of TODD-G, the highest phonon frequency is approximately 56.50 THz, slightly surpassing the 49.11 THz observed in graphene \cite{anees2015temperature,diery2018nature}. The fused 3-8-atom rings in TODD-G have stiff bonds, forbidding atoms to oscillate more freely when contrasted to the graphene case. 

\begin{figure}[!htb]
    \centering
    \includegraphics[width=\linewidth]{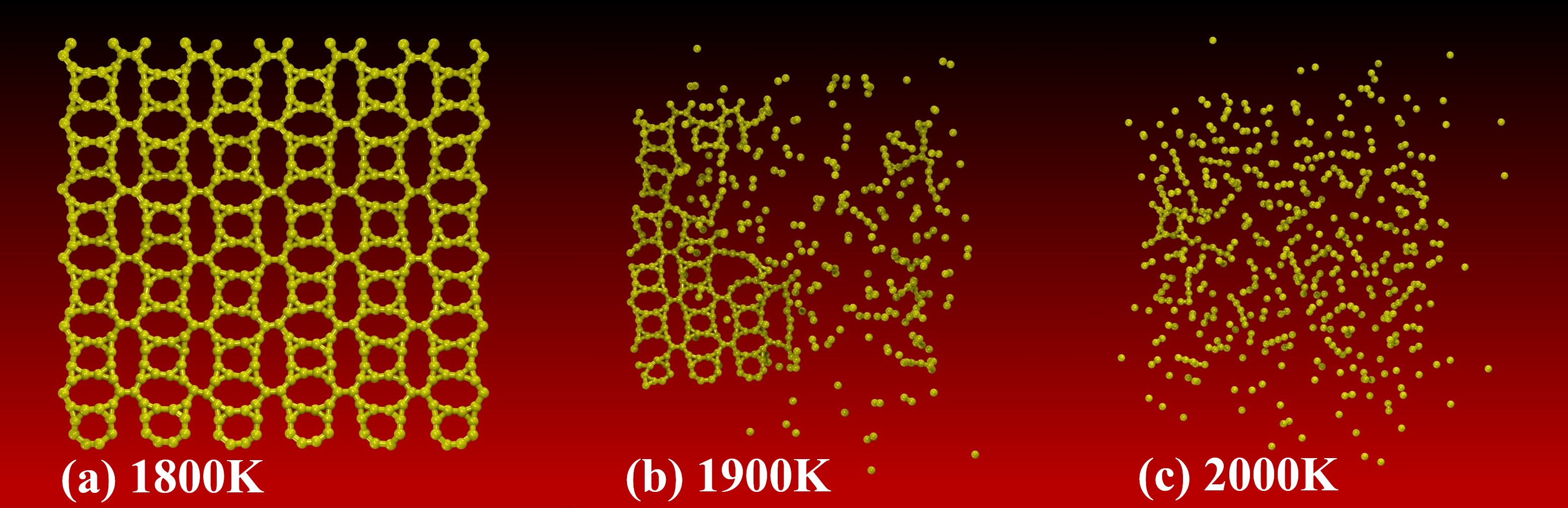}
    \caption{Representative classical reactive MD snapshots of the heating ramp process of TODD-G at different temperatures: 1800 K (a), 1900 K (b), and 2000 K (c) to mark critical temperatures for phase transitions in its topology.}
    \label{fig:heating}
\end{figure}

Figure \ref{fig:heating} illustrates key classical reactive MD snapshots for the dynamic behavior of the TODD-G monolayer throughout the heating simulation. Initially, TODD-G was thermally equilibrated at 300 K while maintaining zero pressure in the $x$ and $y$ directions. This equilibrated structure is still stable at 1800 K, as illustrated in Figure \ref{fig:heating}(a). TODD-G undergoes a significant phase transition around 1900 K, as depicted in Figure \ref{fig:heating}(b) when the system reaches its melting point. We observe the coexistence of linear atomic chains (LACs) and small TODD-G domains in this case. In panel (c) at 2000 K, one can note that the system has melted, resulting in multiple clusters comprising dispersed LACs scattered throughout the simulation box.

\subsection{Electronic and Optical Properties}

\begin{figure}[!htb]
	\centering
	\includegraphics[width=\linewidth]{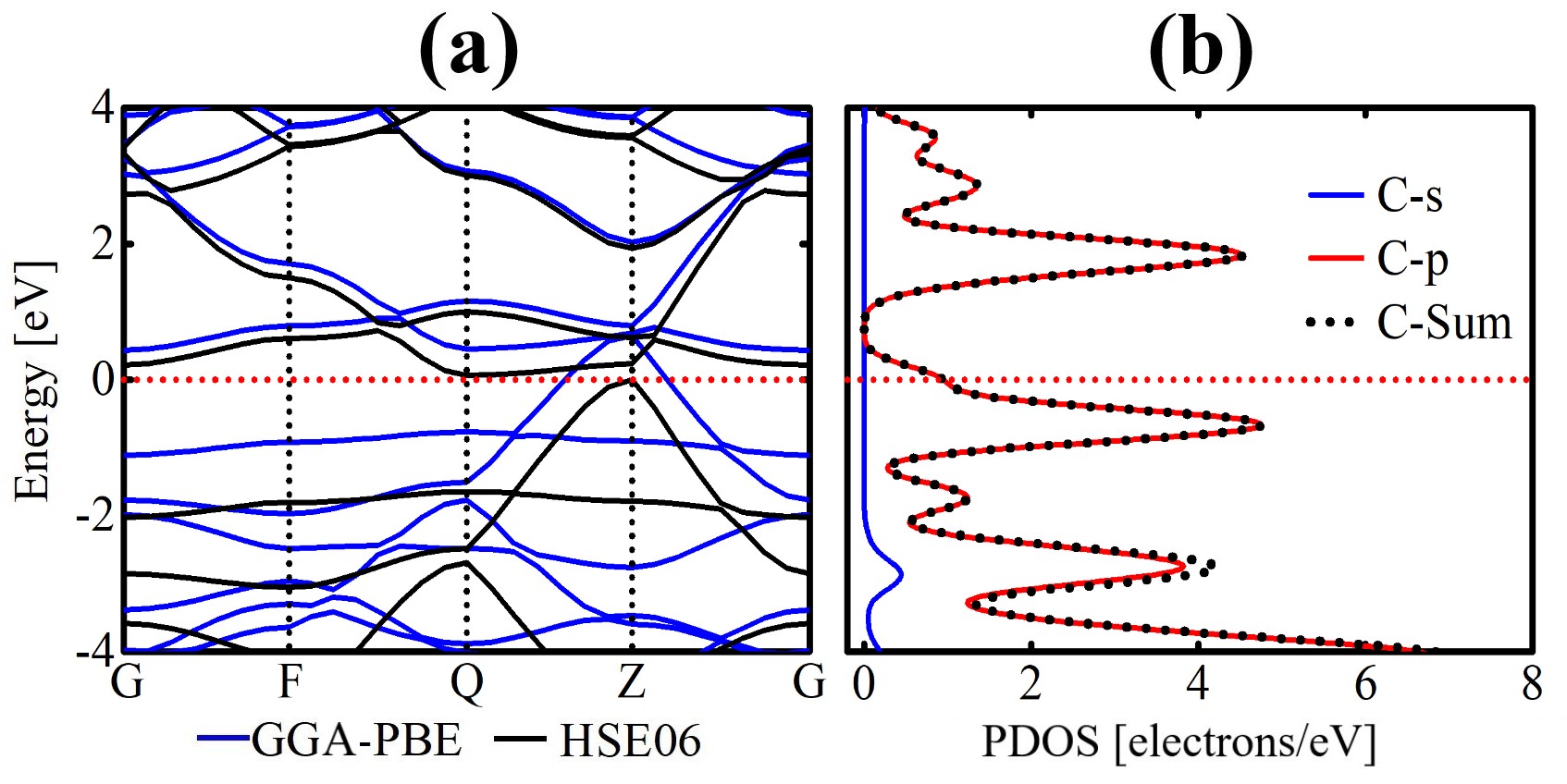}
	\caption{(a) Electronic band structure and (b) partial density of states (PDOS) for TODD-G. The band structure was obtained using the PBE (blue) and HSE06 (black) approaches. PDOS was calculated at the HSE06 level.}
	\label{fig:band}
\end{figure}

Now, we analyze the TODD-G's electronic properties. Figure \ref{fig:band}(a) presents the band structures computed using the PBE and HSE06 methods. Figure \ref{fig:band}(b) provides the PDOS exclusively for the HSE06 method. The band structure obtained with HSE06 showcases a small gap opening of approximately 0.06 meV. In contrast, the PBE-derived band structure doesn't reveal a gap opening. Both approaches point to a metallic signature for TODD-G. It is worth mentioning that PBE calculations often underestimate band gaps. In contrast, HSE06 calculations typically accurately describe a material's electronic and optical properties. Figure \ref{fig:band}(a) further highlights the intrinsic anisotropic conductance of TODD-G. Along the Q-Z direction, it exhibits metallic characteristics while manifesting semiconducting behavior along other paths.   

Figure \ref{fig:band}(b) depicts the PDOS for TODD-G. This figure shows that p-states dominate, suggesting that these states primarily drive electronic transitions and interactions within the material. The p-orbitals are often associated with directional bonding phenomena. In contrast, s-states make a minor contribution to the valence levels. The PDOS results unequivocally establish TODD-G as a metallic material.

Here, we highlight carrier mobility's anisotropic behavior in the TODD-G. Specifically, the carrier mobility along the x-direction significantly surpasses the one in the y-direction, a trend attributed to the smaller effective mass of carriers in the x-direction. The Supplementary Material presents the electron and hole mobility calculation data. The calculated mobilities are 89.25/11.13 and 77.23/10.16 10$^3$ cm$^2$V$^{-1}$s$^{-1}$ for electrons/holes in x and y directions, respectively. TODD-G exhibits enhanced mobility compared to graphene, which can reach 30.0 10$^3$ cm$^2$V$^{-1}$s$^{-1}$ \cite{bolotin2008ultrahigh}. The superior charge carrier mobility observed in the TODD-G compared to graphene can be attributed to their distinct topologies.

\begin{figure}[!htb]
	\centering
	\includegraphics[width=0.6\linewidth]{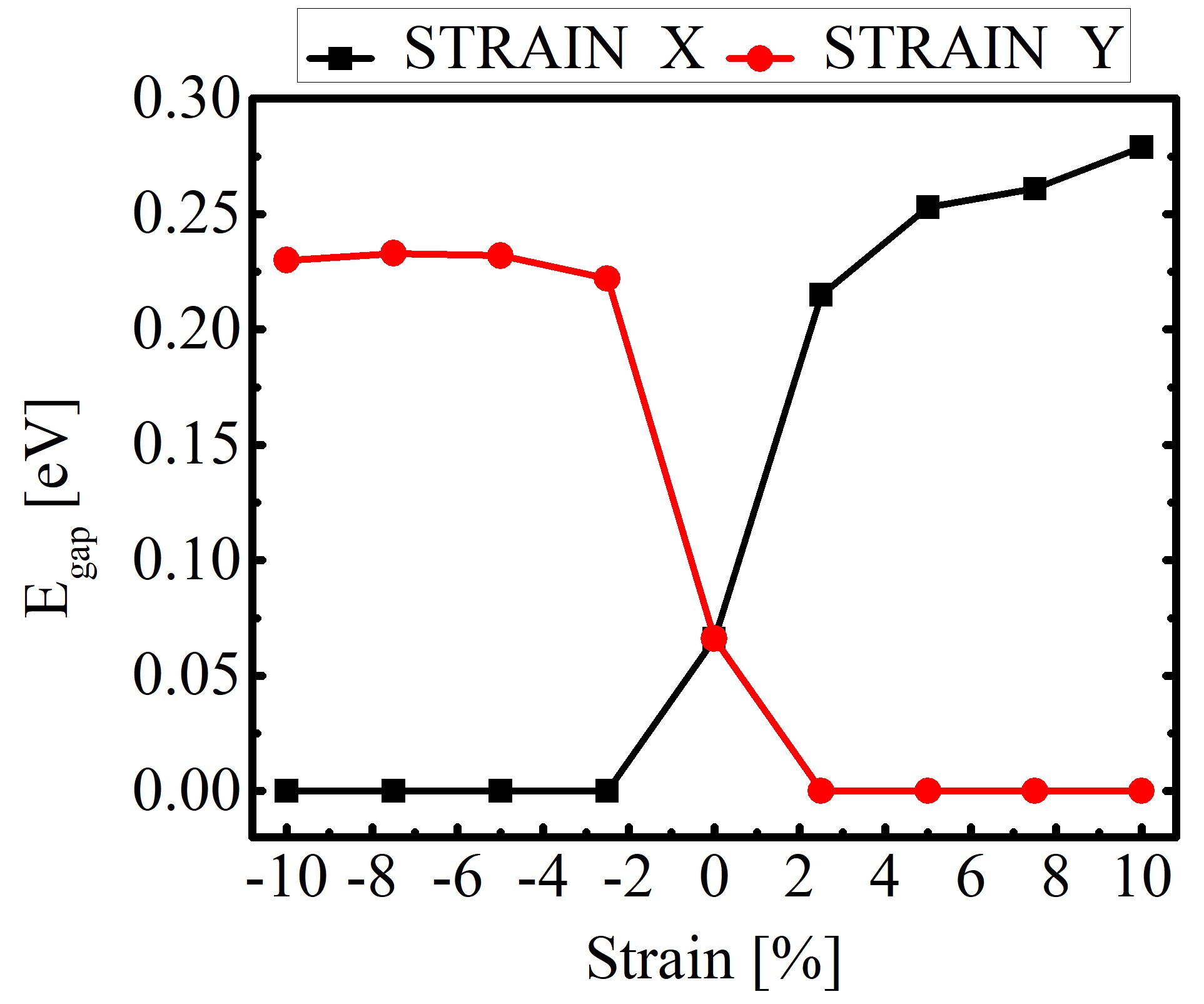}
	\caption{Gap-opening energy as a function of the applied tensile and compressive stress in TOOD-G.}
	\label{fig:strain}
\end{figure}

An unconventional trait was observed in TODD-G regarding its band structure as a response to applied strain. Figure \ref{fig:strain} indicates that the band structure of TODD-G is not easily tunable by strain, as the gap-opening energy ($E_{gap}$) is relatively small, measuring about 0.27 eV for tensile stress in the x-direction and 0.22 eV for compressive strain in the y-direction.

Remarkably, compressive stress in the x-direction and tensile stress in the y-direction result in a complete inversion in the gap-opening trend discussed above, effectively reducing it to zero. This unusual behavior is attributed to the unique topology of TODD-G, which introduces anisotropy in its electronic properties. It is distinct from other materials, like graphene \cite{peng2020strain}, where tensile and compressive stress might more uniformly impact the electronic properties considering directions parallel to the basal plane.  

\begin{figure}[!htb]
	\centering
	\includegraphics[width=\linewidth]{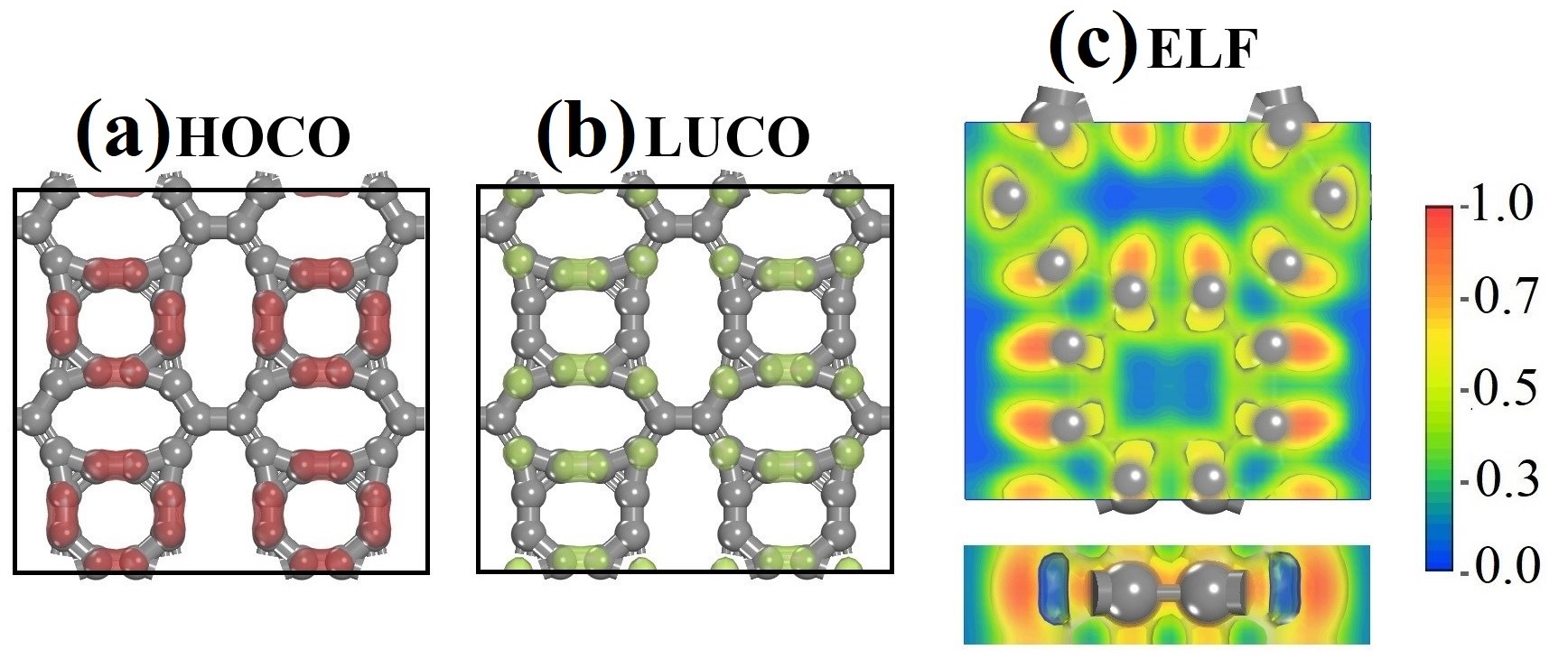}
	\caption{Panel (a) illustrates the highest occupied crystalline orbital (HOCO, in red), and (b) the lowest unoccupied crystalline orbital (LUCO, in green). Panel (c) schematically depicts the electron localization function (ELF).}
	\label{fig:orbitals}
\end{figure}

To gain a deeper understanding of the underlying chemical interactions in TOOD-G, Figure \ref{fig:orbitals} illustrates the highest occupied crystalline orbital (HOCO), the lowest unoccupied crystalline orbital (LUCO), and the electron localization function (ELF). The localization of HOCO and LUCO in TODD-G are shown in Figures \ref{fig:orbitals}(a) and \ref{fig:orbitals}(b), respectively. The HOCO primarily localizes on the 8-membered ring, indicating a higher concentration of low-energy electrons or a greater electron charge density in this region. Simultaneously, LUCO mainly resides on the bonds of the decagons, suggesting a lower density of low-energy electrons or the availability of lower-energy electrons for engaging in chemical interactions in this particular region. This orbital distribution results in a charge imbalance, with the octagonal rings having most of the negative charge.   

In addition to the orbital localization analysis in TODD-G, the ELF, as illustrated in Figure \ref{fig:orbitals}(c), provides valuable insights into the electron distribution within this material. ELF offers a topological perspective on electronic interactions, helping to identify regions with electron localization or delocalization. It assigns a value ranging from 0.0 to 1.0 to each point in space, where values approaching 1.0 indicate strong covalent interactions or lone pair electrons. Conversely, lower values (around $\sim$0.5) suggest delocalization, ionic bonds, or weak Van der Waals interactions. 

The ELF map for TOOD-G highlights a significant electron concentration around the octagonal ring, with values ranging from 0.7 to 1.0. In contrast, bonds between C-C atoms in other lattice regions are characterized by lighter-yellow areas with values near 0.5, indicating a degree of electron delocalization. Materials with valence electrons exhibiting delocalization display metallic-like conductivity, allowing for free-like electron transport. Conversely, materials featuring strong covalent bonds often exhibit semiconductor-like conductivity. In the case of TODD-G, the coexistence of localized and delocalized electrons within its lattice underlies the anisotropic conductance observed in its electronic band structure, as discussed earlier. This complex electron distribution pattern contributes to its unique electronic properties and conductivity behavior.

\begin{figure}[!htb]
	\centering
	\includegraphics[width=\linewidth]{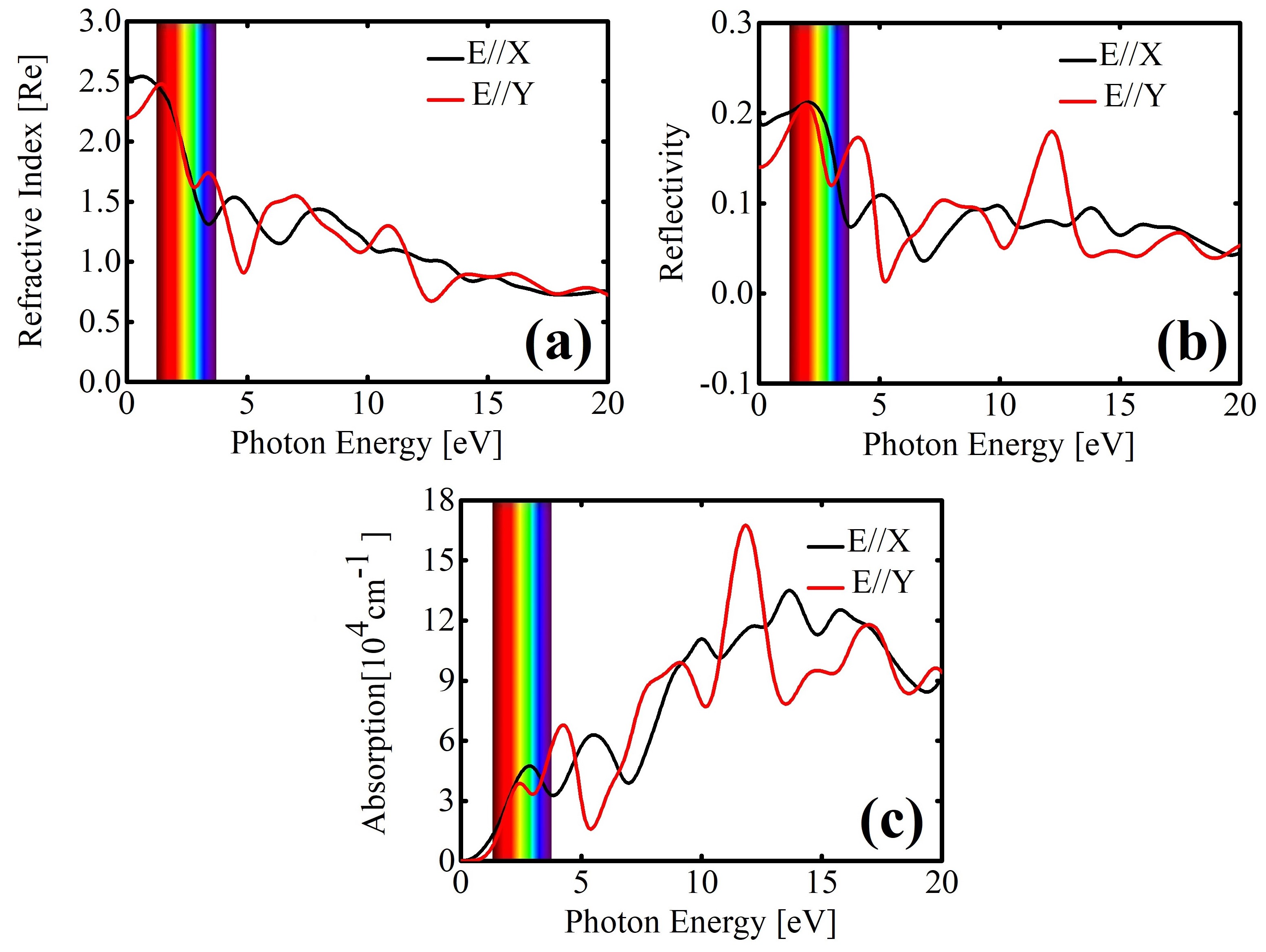}
	\caption{Optical properties as a function of the photon energy for TODD-G: (a) refractive index (a), (b) reflectivity, and (c) absorption coefficient.}
	\label{fig:optical}
\end{figure}

Materials exhibit electronic transitions encompassing interband and intraband processes, crucial determinants of their optical behavior. The lattice topology of TODD-G introduces anisotropy in the underlying optoelectronic properties, endowing distinct optical features along the in-plane polarization directions. In Figure \ref{fig:optical}, we delve into the optical characteristics of TODD-G, with a specific focus on light polarization along the $x$ (E//X) and $y$ (E//Y) directions. Using the HSE06 method for calculations, we explore optical parameters such as the absorption coefficient, refractive index, and reflectivity, all derivable from the complex dielectric function \cite{lima2023dft}.  

Birefringence occurs when the speed of light varies in different polarization directions in the material. In TODD-G, the refractive indices along polarization directions parallel to its basal plane exhibit anisotropy, as depicted in Figure \ref{fig:optical}(a). The most significant refraction occurs at the infrared limit. This observed trend indicates that TODD-G possesses birefringent properties, setting it apart from graphene \cite{rani2014dft}. Moreover, graphene has an intense refraction activity within the Vis-UV ranges \cite{rani2014dft}. There is a decline in the refractive index value, converging to 0.07 for photon energies higher than 18 eV. This convergence indicates that incident UV light is refracted similarly in all directions.   

The reflection function quantifies the ratio of photon energy reflected off a surface to the photon energy incident upon it. In Figure \ref{fig:optical}(b), the TODD-G reflectivity coefficient is presented as a function of photon energy along the $x$ and $y$ directions. Across a range of photon energies from 0 to 20 eV, the reflectivity coefficients remain below 0.03. The maximum reflectivity occurs within the visible region, suggesting efficient transmission of incident light with minimal impact on TODD-G. Reflectivity peaks are observed for photon energies within the range of 10–15 eV, with the reflectivity reaching a maximum value of 0.026 at 1.65 eV. This analysis underscores TODD-G's sensitivity to the reflection of incident Vis-UV light spectrum, similar to graphene \cite{rani2014dft}. Reflectivity peaks that decrease with increasing energy may indicate the existence of electronic resonances or specific vibrations in the material corresponding to these energies. These findings suggest that incident light on TODD-G is predominantly absorbed, meaning its transparency. 

Figure \ref{fig:optical}(c) showcases the spectra of optical absorption coefficients. These coefficients delineate the attenuation of light strength per unit distance traversing the material. As depicted in Figure \ref{fig:optical}(c), TODD-G manifests a high absorption coefficient (10$^{4}$ cm$^{-1}$), attributable to its metallic properties. The first absorption peaks for E//X and E//Y are situated within the visible spectrum, a distinct characteristic compared to observations in graphene \cite{rani2014dft}. TODD-G's initial peak, approximately at 2.9 eV (falling within the red region of the visible spectrum), exhibits a red-shift of about 1.1 eV when compared to graphene, which features its first peak in the UV region (around 4.0 eV) \cite{rani2014dft}. These optical findings suggest that TODD-G holds potential applications as Vis-UV detectors and absorbers.

\subsection{Mechanical Properties}

\begin{figure}[htb!]
	\centering
	\includegraphics[width=\linewidth]{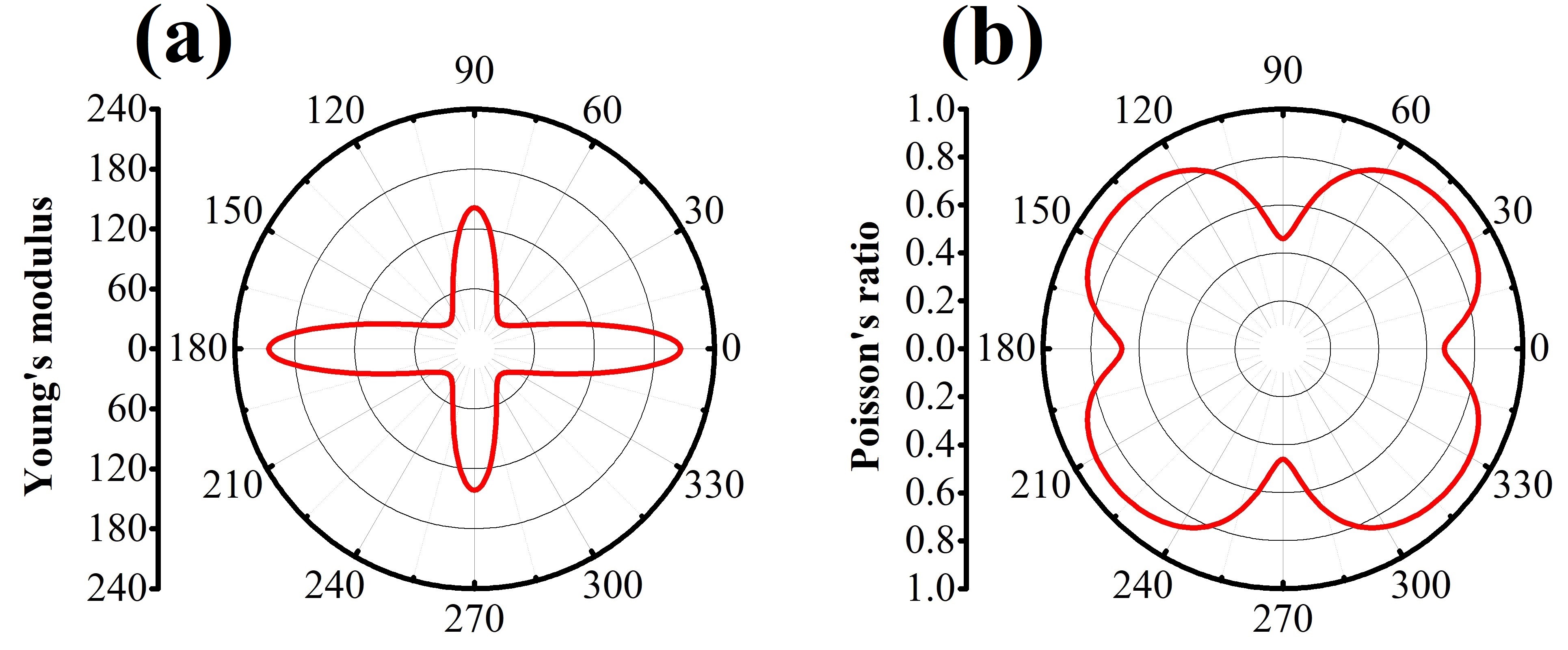}
	\caption{2D representation of (a) Poisson’s ratio and (b) Young's modulus in the xy plane for TODD-G.}
	\label{fig:elastic}
\end{figure}

We now analyze the elastic properties of TODD-G, which play a crucial role in understanding microcracks behavior and its overall durability. To investigate the anisotropy in its mechanical properties, we determine the Poisson's ratio ($\nu(\theta)$) and Young's modulus ($Y(\theta)$) \cite{doi:10.1021/acsami.9b10472,doi:10.1021/acs.jpclett.8b00616} under pressure in the xy plane as:

\begin{equation}
    \displaystyle Y(\theta) = \frac{{C_{11}C_{22} - C_{12}^2}}{{C_{11}\alpha^4 + C_{22}\beta^4 + \left(\frac{{C_{11}C_{22} - C_{12}^2}}{{C_{44}}} - 2C_{12}\right)\alpha^2\beta^2}}
    \label{young}
\end{equation}

\noindent and 

\begin{equation}
    \displaystyle \nu(\theta)= \frac{{(C_{11} + C_{22} - \frac{{C_{11}C_{22} - C_{12}^2}}{{C_{44}}})\alpha^2\beta^2 - C_{12}(\alpha^4 + \beta^4)}}{{C_{11}\alpha^4 + C_{22}\beta^4 + \left(\frac{{C_{11}C_{22} - C_{12}^2}}{{C_{44}}} - 2C_{12}\right)\alpha^2\beta^2}},
    \label{poisson}
\end{equation}

\noindent where, $\alpha=\cos(\theta)$ and $\beta=\sin(\theta)$. The elastic constants of TODD-G are shown in Table \ref{tab:elastic}. Figure \ref{fig:elastic} provides a 2D representation of Young's modulus (Figure \ref{fig:elastic}(a)) and Poisson's ratio (Figure \ref{fig:elastic}(b)) in the xy plane for this material.

\begin{table}[!htp]
\centering
\caption{Elastic constants C$_{ij}$ (GPa) and maximum values for Young's modulus (GPa) ($Y_{MAX}$) and maximum ($\nu_{MAX}$) and ($\nu_{MIN}$) Poisson's ratios.}
\label{tab:elastic}
\begin{tabular}{| l |c|c|c|c|c|c|c|c|}
\hline
 Structure & C$_{11}$ & C$_{12}$ &C$_{22}$ &C$_{44}$ & $Y_{MAX}$  & $\nu_{MAX}$ & $\nu_{MIN}$ \\
 \hline
TODD-G   & $204.57$       & $ 137.47$     & $298.66$     & $9.61$  & $106.20$ & $0.65$ & $0.30$ \\
 \hline
 \end{tabular}
\end{table}

The elastic constants C${11}$, C${22}$, C${12}$, and C${44}$ presented in Table \ref{tab:elastic} satisfy the Born-Huang criteria of the orthorhombic crystal ($C_{11}C_{22} - C_{12}^2>0$ and $C_{44}>0$) \cite{PhysRevB.90.224104,doi:10.1021/acs.jpcc.9b09593}, indicating its good mechanical stability. Additionally, Young's modulus and Poisson's ratio of TODD-G were calculated using Equations \ref{young} and \ref{poisson} (see Figure \ref{fig:elastic}). We found these properties exhibit anisotropic characteristics.

As anticipated, TODD-G exhibits anisotropic behavior in Young's modulus values when subjected to deformation due to its unique ring arrangement within its plane (see Figure \ref{fig:elastic}(a)). The corresponding $Y_{MAX}$ values for deformation in the x and y directions measure approximately 121 GPa and 195 GPa, respectively, which are almost a tenth of the one reported for graphene (1.0 TPa \cite{lee2008measurement}). This disparity can be attributed to the inherent porosity in TODD-G, stemming from the presence of 8-10-12-atom rings and the rigidity of the bonds within fused trigonal rings.

Typical materials fall within the Poisson's ratio range of 0.2 to 0.5 \cite{greaves2011poisson}. A Poisson's ratio of 0.5 characterizes incompressible materials, meaning their lateral dimensions remain almost unchanged under strain. Under uniaxial tensile loading in the x-direction, TODD-G exhibits a maximum Poisson's ratio ($\nu_{MAX}$) of 0.65. This value surpasses that observed in graphene, approximately 0.19 \cite{politano2015probing}. The heightened Poisson ratio in TODD-G is attributed to its lattice arrangement, which features higher porosity than graphene. This increased porosity allows TODD-G to undergo more deformation under tension than graphene, resulting in an elevated Poisson ratio. The intrinsic anisotropy of TODD-G is also evident in Figure \ref{fig:elastic}(b), where the minimum Poisson's ratio ($\nu_{MIN}$) of about 0.30 occurs under strains applied in the y-direction, indicating TODD-G's relative incompressibility in this scenario.

\subsection{Lithium-Ion Adsorption on TODD-G}

\begin{figure}[!htp]
	\centering
	\includegraphics[width=\linewidth]{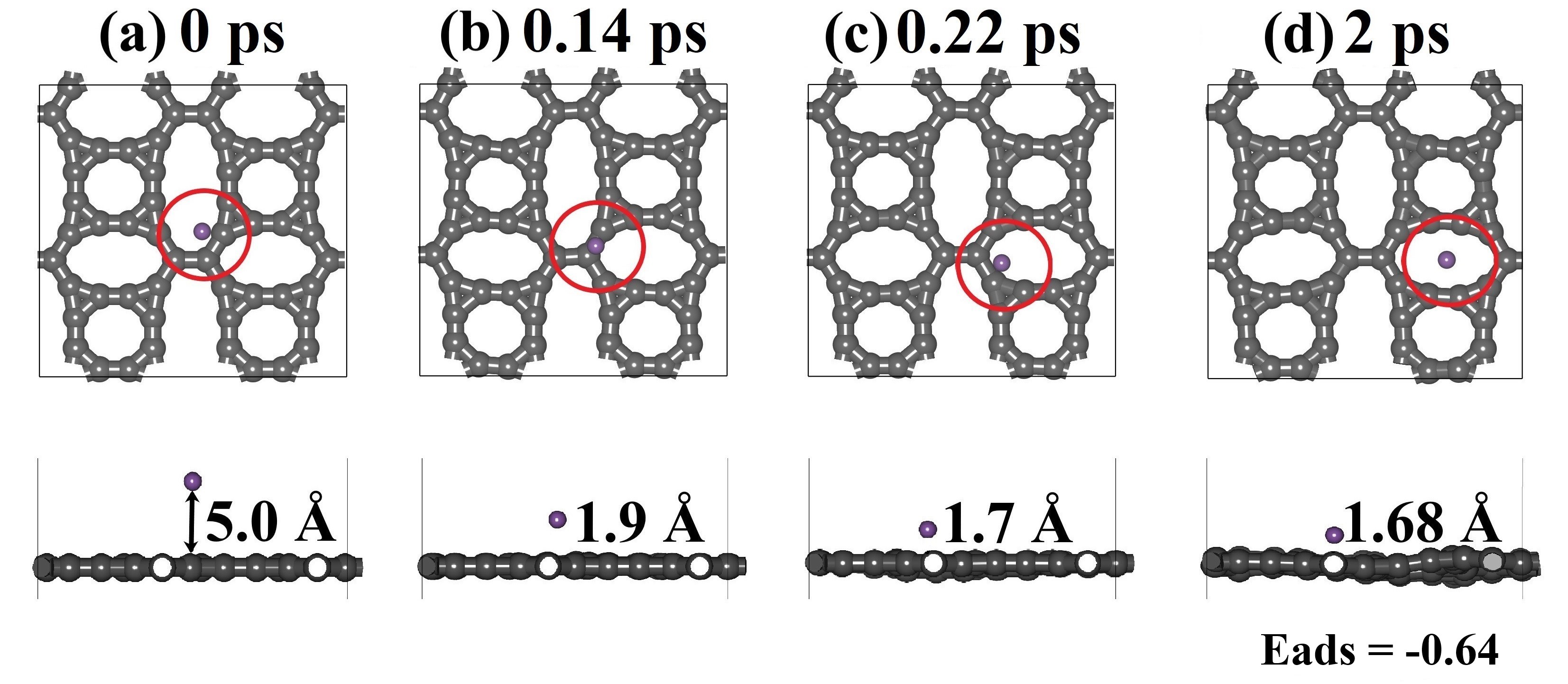}
	\caption{AIMD snapshots for Li adatom diffusion on TODD-G at 500K. This AIMD simulation includes vdW corrections.}
	\label{fig:lidynam}
\end{figure}

Given that Li storage is a primary application focus for 2D carbon-based materials \cite{zhang20162d,kumar2018recent}, we now delve into the Li adsorption process on TODD-G. AIMD simulations, with van der Waals (vdW) corrections within the Grimme scheme \cite{grimme2010consistent,grimme2011effect}, were conducted to investigate the dynamical stability of a system comprising TODD-G and a single Li adatom (refer to Figure \ref{fig:lidynam}). The initial TODD-G/Li system for these simulations was generated using the uncoupled Monte Carlo (UMC) approach facilitated by the Adsorption Locator Modulus, as implemented in Materials Studio \cite{vcerny1985thermodynamical, kirkpatrick1983optimization}.

The UMC approach involves sampling various configurations in the canonical ensemble by creating an extensive set of TODD-G/Li systems, beginning with a trial configuration. These conformations are randomly selected by translating the Li adatom parallel and perpendicular to the basal plane of TODD-G. The quest for the structures with the lowest energy levels is facilitated through the simulated annealing method, with the Metropolis algorithm providing the statistical weights for this process \cite{kirkpatrick1983optimization}. Further details on the UMC approach employed in this study can be found in reference \cite{junior2021adsorption}.

In Figure \ref{fig:lidynam}(a), the AIMD snapshot depicts the initial system at 0 ps. The UMC calculations yield a system with the lowest energy, positioning the Li adatom 5.0 \r{A} above the TODD-G surface. Consistent with prior studies, lithium atoms prefer absorbing on the hollow sites of 2D carbon allotropes \cite{wang2018popgraphene,ferguson2017biphenylene,li2021two}. This pattern is also observed in TODD-G, as illustrated by AIMD snapshots in Figures \ref{fig07}(b-d), showcasing the rapid migration of the Li adatom from the 12-atom ring towards the 10-atom ring within 2 ps. The final Li adsorption energy and distance measure approximately -0.64 eV and 1.68 \r{A}, respectively. Li is very mobile on TODD-G and does not interact as strongly with its surface. Importantly, this adsorption energy is comparable to the values reported for recently predicted 2D carbon allotropes popgraphene (-0.57 to -0.95 eV \cite{wang2018popgraphene}), net-$\tau$ (-0.37 to -0.60 eV \cite{wang2019planar}), and C$_{5678}$ (-0.42 to -0.52 eV \cite{li2021two}). 

These AIMD snapshots highlight that, in TODD-G, the most favorable position for Li adsorption is the hollow region defined by the 10-atom ring, emphasizing a preference for hollow sites over bridge sites. It is worth mentioning that the adsorption energy $E_{ads}$ is defined as $E_{ads}=(E_{system}-E_{TODD-G}-nE_{Li})$, where $E_{system}$ and $E_{TODD-G}$ represent the total energies of the TODD-G structure after and before Li adsorption, and $E_{Li}$ denotes the total energy per Li atom in bulk metal Li. 

\begin{figure}[!htp]
	\centering
	\includegraphics[width=\linewidth]{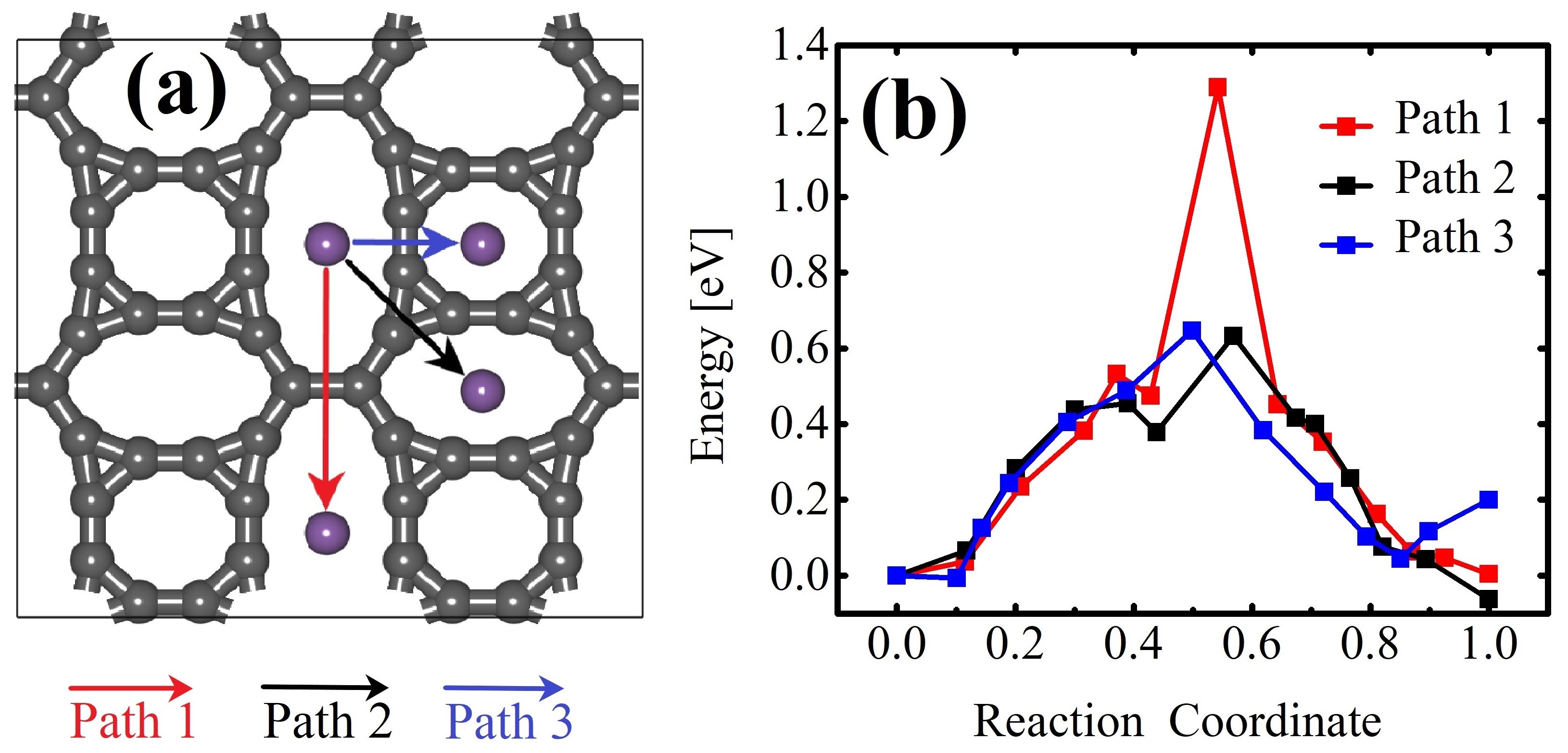}
	\caption{Three possible migration paths (a) and their corresponding diffusion energy profiles (b) for Li diffusion on a TODD-G sheet. These simulations include vdW corrections.}
	\label{fig:barrier}
\end{figure}

To explore lithium mobility on the surface of TODD-G, we selected three potential migration pathways along distinct directions and calculated the migration barrier. The transition states for these migration pathways are depicted in Figure \ref{fig:barrier}(a). The considered pathways are as follows: Pathway 1 from the 12-atom ring to another 12-atom ring (red line), Pathway 2 from the 12-atom ring to the 10-atom ring (black line), and Pathway 3 from the 12-atom ring to the 8-atom ring (blue line).

In examining these pathways, it was observed that the diffusion barrier along the nearest neighboring 12-atom rings (Path 1) is higher, measuring approximately 1.28 eV, as illustrated in Figure \ref{fig:barrier}(b). On the other hand, the diffusion barriers along Paths 2 and 3 are 0.63 eV and 0.64 eV, respectively, surpassing that of graphene (0.31 eV) \cite{guo2015first}, as shown in Figure \ref{fig:barrier}. While the maximum diffusion barrier of TODD-G is higher than that of graphene and $\Psi$-graphene (approximately 0.31 eV) \cite{li2017psi}, it remains comparable to those of $\Theta$-graphene (0.48 eV) \cite{wang2019reconfiguring}, xgraphene (0.49 eV) \cite{wang2019two}, popgraphene (0.55 eV) \cite{wang2018popgraphene}, and C$_{5678}$ (0.44 eV) \cite{li2021two}. It is worth mentioning that it is considerably lower than the biphenylene network (2.44 eV) and phagraphene (2.07 eV) \cite{ferguson2017biphenylene}. The average migration barrier for the Li atom on the surface of TODD-G is 0.85 eV. This low barrier suggests good Li-ion mobility, indicating a favorable charge/discharge rate for TODD-G as an anode material in lithium-ion batteries.

\section{Conclusion}

Through DFT, AIMD, and classical reactive MD simulations, we introduce TODD-G, a novel 2D flat carbon material with a porous topology comprising 3-8-10-12 carbon rings. This structure demonstrates programmable metallic properties designed using a bottom-up approach. TODD-G exhibits a low-energy structure, ensuring dynamic, thermal, and mechanical stability. It has a low average diffusion barrier (about 0.85 eV) and a metallic framework boasting excellent conductivity, emerging as a promising anode material for lithium-ion batteries.

A Dirac cone is above the Fermi level in the TODD-G band structure. The DOS near the Fermi level is formed by 2pz atomic orbitals, confirming its non-magnetic metallic signature. TODD-G has apparent in-plane anisotropic electronic, mechanical, and optical properties, as expected from its topology. Its inherent porosity stems from 8-10-12-atom rings and the rigidity of the bonds within fused trigonal rings, which are responsible for its smaller mechanical resilience than graphene. Classical reactive (ReaxFF) molecular dynamics simulation results suggested its structural integrity with no bond reconstructions at 1800 K.

\section*{Acknowledgements}

This work was financed by the Coordenação de Aperfeiçoamento de Pessoal de Nível Superior (CAPES), Conselho Nacional de Desenvolvimento Cientifico e Tecnológico (CNPq), and Fundação de Apoio à Pesquisa do Distrito Federal (FAP-DF). L.A.R.J. acknowledges the financial support from FAP-DF $00193.00001808/2022-71$ grant and FAPDF-PRONEM grant $00193.00001247/2021-20$, and CNPq grant $350176/2022-1$. L.A.R.J. acknowledges N\'ucleo de Computaç\~ao de Alto Desempenho (N.A.C.A.D.) and for providing the computational facilities. This work used resources of the Centro Nacional de Processamento de Alto Desempenho em São Paulo (CENAPAD-SP). L.A.R.J. and K.A.L.L. also acknowledge CAPES for partially financing this study - Finance Code 88887.691997/2022-00.

\bibliographystyle{unsrt}
\bibliography{references}

\begin{thebibliography}{10}

\bibitem{Novoselov}
K.~S. Novoselov, A.~K. Geim, S.~V. Morozov, D.~Jiang, Y.~Zhang, S.~V. Dubonos, I.~V. Grigorieva, and A.~A. Firsov.
\newblock Electric field effect in atomically thin carbon films.
\newblock {\em Science}, 306(5696):666--669, 2004.

\bibitem{Geim}
A.~K. Geim.
\newblock Graphene: Status and prospects.
\newblock {\em Science}, 324(5934):1530--1534, 2009.

\bibitem{enyashin2011graphene}
Andrey~N Enyashin and Alexander~L Ivanovskii.
\newblock Graphene allotropes.
\newblock {\em physica status solidi (b)}, 248(8):1879--1883, 2011.

\bibitem{wang2015phagraphene}
Zhenhai Wang, Xiang-Feng Zhou, Xiaoming Zhang, Qiang Zhu, Huafeng Dong, Mingwen Zhao, and Artem~R Oganov.
\newblock Phagraphene: a low-energy graphene allotrope composed of 5--6--7 carbon rings with distorted dirac cones.
\newblock {\em Nano letters}, 15(9):6182--6186, 2015.

\bibitem{lu2013two}
Haigang Lu and Si-Dian Li.
\newblock Two-dimensional carbon allotropes from graphene to graphyne.
\newblock {\em Journal of Materials Chemistry C}, 1(23):3677--3680, 2013.

\bibitem{zhang2015penta}
Shunhong Zhang, Jian Zhou, Qian Wang, Xiaoshuang Chen, Yoshiyuki Kawazoe, and Puru Jena.
\newblock Penta-graphene: A new carbon allotrope.
\newblock {\em Proceedings of the National Academy of Sciences}, 112(8):2372--2377, 2015.

\bibitem{terrones2000new}
Humberto Terrones, Mauricio Terrones, E~Hern{\'a}ndez, N~Grobert, Jean-Christophe Charlier, and PM~Ajayan.
\newblock New metallic allotropes of planar and tubular carbon.
\newblock {\em Physical review letters}, 84(8):1716, 2000.

\bibitem{PhysRevB.70.085417}
Savas Berber, Eiji Osawa, and David Tom\'anek.
\newblock Rigid crystalline phases of polymerized fullerenes.
\newblock {\em Phys. Rev. B}, 70:085417, Aug 2004.

\bibitem{wang2018popgraphene}
Shuaiwei Wang, Baocheng Yang, Houyang Chen, and Eli Ruckenstein.
\newblock Popgraphene: a new 2d planar carbon allotrope composed of 5--8--5 carbon rings for high-performance lithium-ion battery anodes from bottom-up programming.
\newblock {\em Journal of Materials Chemistry A}, 6(16):6815--6821, 2018.

\bibitem{tromer2023mechanical}
Raphael~M Tromer, Marcelo~L Pereira~J\'unior, Kleuton~A L.~Lima, Alexandre~F Fonseca, Luciano~R da~Silva, Douglas~S Galva\~ao, and Luiz~A Ribeiro~Junior.
\newblock Mechanical, electronic, and optical properties of 8-16-4 graphyne: A 2d carbon allotrope with dirac cones.
\newblock {\em The Journal of Physical Chemistry C}, 2023.

\bibitem{junior2023irida}
ML~Pereira J{\'u}nior, Wiliam~Ferreira da~Cunha, William~Ferreira Giozza, Rafael~Timoteo de~Sousa~Junior, and LA~Ribeiro Junior.
\newblock Irida-graphene: A new 2d carbon allotrope.
\newblock {\em FlatChem}, 37:100469, 2023.

\bibitem{Qitang}
Qitang Fan, Linghao Yan, Matthias~W. Tripp, Ondřej Krejčí, Stavrina Dimosthenous, Stefan~R. Kachel, Mengyi Chen, Adam~S. Foster, Ulrich Koert, Peter Liljeroth, and J.~Michael Gottfried.
\newblock Biphenylene network: A nonbenzenoid carbon allotrope.
\newblock {\em Science}, 372(6544):852--856, 2021.

\bibitem{Desyatkin}
Victor~G. Desyatkin, William~B. Martin, Ali~E. Aliev, Nathaniel~E. Chapman, Alexandre~F. Fonseca, Douglas~S. Galvão, Ericka~Roy Miller, Kevin~H. Stone, Zhong Wang, Dante Zakhidov, F.~Ted Limpoco, Sarah~R. Almahdali, Shane~M. Parker, Ray~H. Baughman, and Valentin~O. Rodionov.
\newblock Scalable synthesis and characterization of multilayer $\gamma$-graphyne, new carbon crystals with a small direct band gap.
\newblock {\em Journal of the American Chemical Society}, 144(39):17999--18008, 2022.
\newblock PMID: 36130080.

\bibitem{hou2022synthesis}
Lingxiang Hou, Xueping Cui, Bo~Guan, Shaozhi Wang, Ruian Li, Yunqi Liu, Daoben Zhu, and Jian Zheng.
\newblock Synthesis of a monolayer fullerene network.
\newblock {\em Nature}, 606(7914):507--510, 2022.

\bibitem{zheng2015two}
Xiaoyu Zheng, Jiayan Luo, Wei Lv, Da-Wei Wang, and Quan-Hong Yang.
\newblock Two-dimensional porous carbon: Synthesis and ion-transport properties.
\newblock {\em Advanced Materials}, 27(36):5388--5395, 2015.

\bibitem{borchardt2017toward}
Lars Borchardt, Qi-Long Zhu, Mirian~E Casco, Reinhard Berger, Xiaodong Zhuang, Stefan Kaskel, Xinliang Feng, and Qiang Xu.
\newblock Toward a molecular design of porous carbon materials.
\newblock {\em Materials Today}, 20(10):592--610, 2017.

\bibitem{tao2020advanced}
You Tao, Zhu-Yin Sui, and Bao-Hang Han.
\newblock Advanced porous graphene materials: From in-plane pore generation to energy storage applications.
\newblock {\em Journal of Materials Chemistry A}, 8(13):6125--6143, 2020.

\bibitem{fang20222d}
Yan Fang, Yuxin Liu, Lu~Qi, Yurui Xue, and Yuliang Li.
\newblock 2d graphdiyne: an emerging carbon material.
\newblock {\em Chemical Society Reviews}, 51(7):2681--2709, 2022.

\bibitem{zhang2013controlling}
Long Zhang, Xi~Yang, Fan Zhang, Guankui Long, Tengfei Zhang, Kai Leng, Yawei Zhang, Yi~Huang, Yanfeng Ma, Mingtao Zhang, et~al.
\newblock Controlling the effective surface area and pore size distribution of sp2 carbon materials and their impact on the capacitance performance of these materials.
\newblock {\em Journal of the American Chemical Society}, 135(15):5921--5929, 2013.

\bibitem{macha20192d}
Michal Macha, Sanjin Marion, Vishal~VR Nandigana, and Aleksandra Radenovic.
\newblock 2d materials as an emerging platform for nanopore-based power generation.
\newblock {\em Nature Reviews Materials}, 4(9):588--605, 2019.

\bibitem{yao2018scalable}
Lei Yao, Qin Wu, Peixin Zhang, Junmin Zhang, Dongrui Wang, Yongliang Li, Xiangzhong Ren, Hongwei Mi, Libo Deng, and Zijian Zheng.
\newblock Scalable 2d hierarchical porous carbon nanosheets for flexible supercapacitors with ultrahigh energy density.
\newblock {\em Advanced materials}, 30(11):1706054, 2018.

\bibitem{kochaev2023ionic}
A~Kochaev, K~Katin, and M~Maslov.
\newblock On ionic transport through pores in a borophene--graphene membrane.
\newblock {\em Materials Today Chemistry}, 30:101512, 2023.

\bibitem{dolina2023thermal}
Ekaterina~S Dolina, Pavel~A Kulyamin, Anastasiya~A Grekova, Alexey~I Kochaev, Mikhail~M Maslov, and Konstantin~P Katin.
\newblock Thermal stability and vibrational properties of the 6, 6, 12-graphyne-based isolated molecules and two-dimensional crystal.
\newblock {\em Materials}, 16(5):1964, 2023.

\bibitem{mortazavi2023electronic}
Bohayra Mortazavi.
\newblock Electronic, thermal and mechanical properties of carbon and boron nitride holey graphyne monolayers.
\newblock {\em Materials}, 16(20):6642, 2023.

\bibitem{mortazavi2023theoretical}
Bohayra Mortazavi.
\newblock A theoretical investigation of the structural, electronic and mechanical properties of pristine and nitrogen-terminated carbon nanoribbons composed of 4--5--6--8-membered rings.
\newblock {\em Journal of Composites Science}, 7(7):269, 2023.

\bibitem{Yu}
Yang-Xin Yu.
\newblock Graphenylene: a promising anode material for lithium-ion batteries with high mobility and storage.
\newblock {\em Journal of Materials Chemistry A}, 1(43):13559--13566, 2013.

\bibitem{kim2018study}
Jaewook Kim, Sungwoo Kang, Jaechang Lim, and Woo~Youn Kim.
\newblock Study of li adsorption on graphdiyne using hybrid dft calculations.
\newblock {\em ACS applied materials \& interfaces}, 11(3):2677--2683, 2018.

\bibitem{li}
Xiaoyin Li, Qian Wang, and Puru Jena.
\newblock $\psi$-graphene: a new metallic allotrope of planar carbon with potential applications as anode materials for lithium-ion batteries.
\newblock {\em The journal of physical chemistry letters}, 8(14):3234--3241, 2017.

\bibitem{wang2019planar}
Xiao Wang, Zhihao Feng, Ju~Rong, Yannan Zhang, Yi~Zhong, Jing Feng, Xiaohua Yu, and Zhaolin Zhan.
\newblock Planar net-$\tau$: A new high-performance metallic carbon anode material for lithium-ion batteries.
\newblock {\em Carbon}, 142:438--444, 2019.

\bibitem{Da}
Da~Li.
\newblock Two-dimensional c 5678: a promising carbon-based high-performance lithium-ion battery anode.
\newblock {\em Materials Advances}, 2(1):398--402, 2021.

\bibitem{zhang2023li}
Ya-Fei Zhang and Junxiong Guo.
\newblock Li-decorated 2d irida-graphene as a potential hydrogen storage material: A dispersion-corrected density functional theory calculations.
\newblock {\em International Journal of Hydrogen Energy}, 2023.

\bibitem{mortazavi2022electronic}
Bohayra Mortazavi, Fazel Shojaei, Masoud Shahrokhi, Timon Rabczuk, Alexander~V Shapeev, and Xiaoying Zhuang.
\newblock Electronic, optical, mechanical and li-ion storage properties of novel benzotrithiophene-based graphdiyne monolayers explored by first principles and machine learning.
\newblock {\em Batteries}, 8(10):194, 2022.

\bibitem{peng2020strain}
Zhiwei Peng, Xiaolin Chen, Yulong Fan, David~J Srolovitz, and Dangyuan Lei.
\newblock Strain engineering of 2d semiconductors and graphene: from strain fields to band-structure tuning and photonic applications.
\newblock {\em Light: Science \& Applications}, 9(1):190, 2020.

\bibitem{heine2015transition}
Thomas Heine.
\newblock Transition metal chalcogenides: ultrathin inorganic materials with tunable electronic properties.
\newblock {\em Accounts of chemical research}, 48(1):65--72, 2015.

\bibitem{chaves2020bandgap}
A~Chaves, Javad~G Azadani, Hussain Alsalman, DR~Da~Costa, R~Frisenda, AJ~Chaves, Seung~Hyun Song, Young~Duck Kim, Daowei He, Jiadong Zhou, et~al.
\newblock Bandgap engineering of two-dimensional semiconductor materials.
\newblock {\em npj 2D Materials and Applications}, 4(1):29, 2020.

\bibitem{zeng2015band}
Qingsheng Zeng, Hong Wang, Wei Fu, Yongji Gong, Wu~Zhou, Pulickel~M Ajayan, Jun Lou, and Zheng Liu.
\newblock Band engineering for novel two-dimensional atomic layers.
\newblock {\em Small}, 11(16):1868--1884, 2015.

\bibitem{kulyamin2023electronic}
Pavel~A Kulyamin, Aleksey~I Kochaev, Mikhail~M Maslov, Roberto Flores-Moreno, Savas Kaya, and Konstantin~P Katin.
\newblock Electronic and optical characteristics of graphene on the molybdenum ditelluride substrate under the uniform mechanical stress.
\newblock {\em Diamond and Related Materials}, page 110547, 2023.

\bibitem{ClarkSegal}
Stewart~J. Clark, Matthew~D. Segall, Chris~J. Pickard, Phil~J. Hasnip, Matt I.~J. Probert, Keith Refson, and Mike~C. Payne.
\newblock First principles methods using castep.
\newblock {\em Zeitschrift für Kristallographie - Crystalline Materials}, 220(5-6):567--570, 2005.

\bibitem{systemes2017biovia}
Dassault Syst{\`e}mes.
\newblock Biovia materials studio.
\newblock {\em San Diego}, 2017.

\bibitem{perdew1996generalized}
John~P Perdew, Kieron Burke, and Matthias Ernzerhof.
\newblock Generalized gradient approximation made simple.
\newblock {\em Physical review letters}, 77(18):3865, 1996.

\bibitem{heyd2003hybrid}
Jochen Heyd, Gustavo~E Scuseria, and Matthias Ernzerhof.
\newblock Hybrid functionals based on a screened coulomb potential.
\newblock {\em The Journal of chemical physics}, 118(18):8207--8215, 2003.

\bibitem{HEAD1985264}
John~D. Head and Michael~C. Zerner.
\newblock A broyden—fletcher—goldfarb—shanno optimization procedure for molecular geometries.
\newblock {\em Chemical Physics Letters}, 122(3):264--270, 1985.

\bibitem{PFROMMER1997233}
Bernd~G. Pfrommer, Michel Côté, Steven~G. Louie, and Marvin~L. Cohen.
\newblock Relaxation of crystals with the quasi-newton method.
\newblock {\em Journal of Computational Physics}, 131(1):233--240, 1997.

\bibitem{PhysRevLett.45.566}
D.~M. Ceperley and B.~J. Alder.
\newblock Ground state of the electron gas by a stochastic method.
\newblock {\em Phys. Rev. Lett.}, 45:566--569, Aug 1980.

\bibitem{PhysRevB.23.5048}
J.~P. Perdew and Alex Zunger.
\newblock Self-interaction correction to density-functional approximations for many-electron systems.
\newblock {\em Phys. Rev. B}, 23:5048--5079, May 1981.

\bibitem{zuo1992elastic}
LIANG Zuo, MICHEL Humbert, and CLAUDE Esling.
\newblock Elastic properties of polycrystals in the voigt-reuss-hill approximation.
\newblock {\em Journal of applied crystallography}, 25(6):751--755, 1992.

\bibitem{chung1967voigt}
DH~Chung and WR~Buessem.
\newblock The voigt-reuss-hill approximation and elastic moduli of polycrystalline mgo, caf2, $\beta$-zns, znse, and cdte.
\newblock {\em Journal of Applied Physics}, 38(6):2535--2540, 1967.

\bibitem{nose1984unified}
Shuichi Nos{\'e}.
\newblock A unified formulation of the constant temperature molecular dynamics methods.
\newblock {\em The Journal of chemical physics}, 81(1):511--519, 1984.

\bibitem{sangiovanni2018ab}
Davide~Giuseppe Sangiovanni, GK~Gueorguiev, and Anelia Kakanakova-Georgieva.
\newblock Ab initio molecular dynamics of atomic-scale surface reactions: Insights into metal organic chemical vapor deposition of aln on graphene.
\newblock {\em Physical Chemistry Chemical Physics}, 20(26):17751--17761, 2018.

\bibitem{lundgren2022perspective}
Christoffer Lundgren, Anelia Kakanakova-Georgieva, and Gueorgui~K Gueorguiev.
\newblock A perspective on thermal stability and mechanical properties of 2d indium bismide from ab initio molecular dynamics.
\newblock {\em Nanotechnology}, 33(33):335706, 2022.

\bibitem{lima2023dft}
KA~Lopes Lima and LA~Ribeiro Junior.
\newblock A dft study on the mechanical, electronic, thermodynamic, and optical properties of gan and aln counterparts of biphenylene network.
\newblock {\em Materials Today Communications}, 37:107183, 2023.

\bibitem{van2001reaxff}
Adri~CT Van~Duin, Siddharth Dasgupta, Francois Lorant, and William~A Goddard.
\newblock Reaxff: a reactive force field for hydrocarbons.
\newblock {\em The Journal of Physical Chemistry A}, 105(41):9396--9409, 2001.

\bibitem{mueller2010development}
Jonathan~E Mueller, Adri~CT van Duin, and William~A Goddard~III.
\newblock Development and validation of reaxff reactive force field for hydrocarbon chemistry catalyzed by nickel.
\newblock {\em The Journal of Physical Chemistry C}, 114(11):4939--4949, 2010.

\bibitem{evans1985nose}
Denis~J Evans and Brad~Lee Holian.
\newblock The nose--hoover thermostat.
\newblock {\em The Journal of chemical physics}, 83(8):4069--4074, 1985.

\bibitem{enyashin}
Andrey~N Enyashin and Alexander~L Ivanovskii.
\newblock Graphene allotropes.
\newblock {\em physica status solidi (b)}, 248(8):1879--1883, 2011.

\bibitem{Heimann1997CarbonAA}
Robert~B. Heimann, S.~E. Evsvukov, and Yoshinori Koga.
\newblock Carbon allotropes: a suggested classification scheme based on valence orbital hybridization.
\newblock {\em Carbon}, 35:1654--1658, 1997.

\bibitem{peng2012mechanical}
Qing Peng, Wei Ji, and Suvranu De.
\newblock Mechanical properties of graphyne monolayers: a first-principles study.
\newblock {\em Physical Chemistry Chemical Physics}, 14(38):13385--13391, 2012.

\bibitem{valencia2006lithium}
Felipe Valencia, Aldo~H Romero, Francesco Ancilotto, and Pier~Luigi Silvestrelli.
\newblock Lithium adsorption on graphite from density functional theory calculations.
\newblock {\em The Journal of Physical Chemistry B}, 110(30):14832--14841, 2006.

\bibitem{PhysRevLett}
Mark~T. Lusk and L.~D. Carr.
\newblock Nanoengineering defect structures on graphene.
\newblock {\em Phys. Rev. Lett.}, 100:175503, Apr 2008.

\bibitem{wang2019dhq}
Xiao Wang, Li~Chen, Zhentao Yuan, Ju~Rong, Jing Feng, Iqbal Muzammil, Xiaohua Yu, Yannan Zhang, and Zhaolin Zhan.
\newblock Dhq-graphene: a novel two-dimensional defective graphene for corrosion-resistant coating.
\newblock {\em Journal of Materials Chemistry A}, 7(15):8967--8974, 2019.

\bibitem{PhysRevB}
Hongki Min, Bhagawan Sahu, Sanjay~K. Banerjee, and A.~H. MacDonald.
\newblock Ab initio theory of gate induced gaps in graphene bilayers.
\newblock {\em Phys. Rev. B}, 75:155115, Apr 2007.

\bibitem{anees2015temperature}
P~Anees, MC~Valsakumar, and BK~Panigrahi.
\newblock Temperature dependent phonon frequency shift and structural stability of free-standing graphene: a spectral energy density analysis.
\newblock {\em 2D Materials}, 2(3):035014, 2015.

\bibitem{diery2018nature}
WA~Diery, Elie~A Moujaes, and RW~Nunes.
\newblock Nature of localized phonon modes of tilt grain boundaries in graphene.
\newblock {\em Carbon}, 140:250--258, 2018.

\bibitem{bolotin2008ultrahigh}
Kirill~I Bolotin, KJ~Sikes, Zhifang Jiang, M~Klima, G~Fudenberg, James Hone, Phaly Kim, and Horst~L Stormer.
\newblock Ultrahigh electron mobility in suspended graphene.
\newblock {\em Solid state communications}, 146(9-10):351--355, 2008.

\bibitem{rani2014dft}
Pooja Rani, Girija~S Dubey, and VK~Jindal.
\newblock Dft study of optical properties of pure and doped graphene.
\newblock {\em Physica E: Low-dimensional Systems and Nanostructures}, 62:28--35, 2014.

\bibitem{doi:10.1021/acsami.9b10472}
Bing Wang, Qisheng Wu, Yehui Zhang, Liang Ma, and Jinlan Wang.
\newblock Auxetic b4n monolayer: A promising 2d material with in-plane negative poisson’s ratio and large anisotropic mechanics.
\newblock {\em ACS Applied Materials \& Interfaces}, 11(36):33231--33237, 2019.

\bibitem{doi:10.1021/acs.jpclett.8b00616}
Yu~Zhao, Xiaoyin Li, Junyi Liu, Cunzhi Zhang, and Qian Wang.
\newblock A new anisotropic dirac cone material: A b2s honeycomb monolayer.
\newblock {\em The Journal of Physical Chemistry Letters}, 9(7):1815--1820, 2018.

\bibitem{PhysRevB.90.224104}
F\'elix Mouhat and Fran\ifmmode \mbox{\c{c}}\else \c{c}\fi{}ois-Xavier Coudert.
\newblock Necessary and sufficient elastic stability conditions in various crystal systems.
\newblock {\em Phys. Rev. B}, 90:224104, Dec 2014.

\bibitem{doi:10.1021/acs.jpcc.9b09593}
Yiran Ying, Ke~Fan, Sicong Zhu, Xin Luo, and Haitao Huang.
\newblock Theoretical investigation of monolayer rhtecl semiconductors as photocatalysts for water splitting.
\newblock {\em The Journal of Physical Chemistry C}, 124(1):639--646, 2020.

\bibitem{lee2008measurement}
Changgu Lee, Xiaoding Wei, Jeffrey~W Kysar, and James Hone.
\newblock Measurement of the elastic properties and intrinsic strength of monolayer graphene.
\newblock {\em science}, 321(5887):385--388, 2008.

\bibitem{greaves2011poisson}
George~Neville Greaves, A~Lindsay Greer, Roderic~S Lakes, and Tanguy Rouxel.
\newblock Poisson's ratio and modern materials.
\newblock {\em Nature materials}, 10(11):823--837, 2011.

\bibitem{politano2015probing}
Antonio Politano and Gennaro Chiarello.
\newblock Probing the young’s modulus and poisson’s ratio in graphene/metal interfaces and graphite: a comparative study.
\newblock {\em Nano Research}, 8:1847--1856, 2015.

\bibitem{zhang20162d}
Xiaoyan Zhang, Lili Hou, Artur Ciesielski, and Paolo Samor{\`\i}.
\newblock 2d materials beyond graphene for high-performance energy storage applications.
\newblock {\em Advanced Energy Materials}, 6(23):1600671, 2016.

\bibitem{kumar2018recent}
Rajesh Kumar, Ednan Joanni, Rajesh~K Singh, Dinesh~P Singh, and Stanislav~A Moshkalev.
\newblock Recent advances in the synthesis and modification of carbon-based 2d materials for application in energy conversion and storage.
\newblock {\em Progress in Energy and Combustion Science}, 67:115--157, 2018.

\bibitem{grimme2010consistent}
Stefan Grimme, Jens Antony, Stephan Ehrlich, and Helge Krieg.
\newblock A consistent and accurate ab initio parametrization of density functional dispersion correction (dft-d) for the 94 elements h-pu.
\newblock {\em The Journal of chemical physics}, 132(15), 2010.

\bibitem{grimme2011effect}
Stefan Grimme, Stephan Ehrlich, and Lars Goerigk.
\newblock Effect of the damping function in dispersion corrected density functional theory.
\newblock {\em Journal of computational chemistry}, 32(7):1456--1465, 2011.

\bibitem{vcerny1985thermodynamical}
Vladim{\'\i}r {\v{C}}ern{\`y}.
\newblock Thermodynamical approach to the traveling salesman problem: An efficient simulation algorithm.
\newblock {\em Journal of optimization theory and applications}, 45:41--51, 1985.

\bibitem{kirkpatrick1983optimization}
Scott Kirkpatrick, C~Daniel Gelatt~Jr, and Mario~P Vecchi.
\newblock Optimization by simulated annealing.
\newblock {\em science}, 220(4598):671--680, 1983.

\bibitem{junior2021adsorption}
Luiz A~Ribeiro Junior, Raphael~M Tromer, Ramiro~M Dos~Santos, and Douglas~S Galvao.
\newblock On the adsorption mechanism of caffeine on mapbi 3 perovskite surfaces: a combined umc-dft study.
\newblock {\em Physical Chemistry Chemical Physics}, 23(18):10807--10813, 2021.

\bibitem{ferguson2017biphenylene}
David Ferguson, Debra~J Searles, and Marlies Hankel.
\newblock Biphenylene and phagraphene as lithium ion battery anode materials.
\newblock {\em ACS applied materials \& interfaces}, 9(24):20577--20584, 2017.

\bibitem{li2021two}
Da~Li.
\newblock Two-dimensional c 5678: a promising carbon-based high-performance lithium-ion battery anode.
\newblock {\em Materials Advances}, 2(1):398--402, 2021.

\bibitem{guo2015first}
Gen-Cai Guo, Da~Wang, Xiao-Lin Wei, Qi~Zhang, Hao Liu, Woon-Ming Lau, and Li-Min Liu.
\newblock First-principles study of phosphorene and graphene heterostructure as anode materials for rechargeable li batteries.
\newblock {\em The journal of physical chemistry letters}, 6(24):5002--5008, 2015.

\bibitem{li2017psi}
Xiaoyin Li, Qian Wang, and Puru Jena.
\newblock $\psi$-graphene: a new metallic allotrope of planar carbon with potential applications as anode materials for lithium-ion batteries.
\newblock {\em The journal of physical chemistry letters}, 8(14):3234--3241, 2017.

\bibitem{wang2019reconfiguring}
Shuaiwei Wang, Baocheng Yang, Houyang Chen, and Eli Ruckenstein.
\newblock Reconfiguring graphene for high-performance metal-ion battery anodes.
\newblock {\em Energy Storage Materials}, 16:619--624, 2019.

\bibitem{wang2019two}
Shuaiwei Wang, Yubing Si, Baocheng Yang, Eli Ruckenstein, and Houyang Chen.
\newblock Two-dimensional carbon-based auxetic materials for broad-spectrum metal-ion battery anodes.
\newblock {\em The journal of physical chemistry letters}, 10(12):3269--3275, 2019.

\end{thebibliography}



\section*{Supplementary material}

# Supplementary Information for
# TODD-Graphene: A Novel Porous 2D Carbon Allotrope for
# High-Performance Lithium-Ion Batteries

E. J. A. dos Santos,[1,2] K. A. Lopes Lima,[1,2] and L. A. Ribeiro Junior[1,2]

[1]*Institute of Physics, University of Brasília, 70910-900, Brasília, Brazil*

[2]*Computational Materials Laboratory, LCCMat, Institute of Physics,
University of Brasília, 70910-900, Brasília, Brazil*

## I. ELECTRON AND HOLE MOBILITY IN TODD-G

To gain a thorough insight into the electronic conductance in TODD-G, we have expanded our examination to consider charge carrier mobility in this material. The analysis uses the deformation potential (DP) theory [1], a well-established approach based on the effective mass approximation []. This method is widely employed for assessing carrier mobility in 2D layered semiconductors. In this way, the carrier mobility ($\mu$) is defined as follows:

$$\mu = \frac{e\hbar C_{2D}}{k_B T m_i^* m_d (E_1)^2}. \quad (1)$$

In the expression for the average effective mass, $m_d = \sqrt{m_x^* m_y^*}$. Here, $C_{2D}$ denotes the in-plane stiffness, defined as $C_{2D} = \left(\partial^2 E/\partial \delta^2\right)/S_0$, where $E$, $\delta$, and $S_0$ represent the total energy, applied strain, and the area of the system, respectively.

The effective masses for electrons ($e$) and holes ($h$) are denoted by $m_i^*$. This parameter is calculated by fitting the band dispersion to

$$m^* = \hbar^2 \left(\frac{\partial^2 E(k)}{\partial k^2}\right)^{-1}. \quad (2)$$

Additionally, $E_1$ is calculated as $dE_{edge}/d\delta$, where $\delta$ represents the applied strain in increments of 0.5%. The term $E_{edge}$ corresponds to the energy of the band edges, specifically the valence band maximum (VBM) for holes and the conduction band minimum (CBM) for electrons. The positions of VBM and CBM are identified at the G and Z points. Furthermore, in the context of the equation, $k_B$, $T$, $e$, and $\hbar$ denote Boltzmann's constant, temperature, the elementary charge of an electron, and Planck's constant, respectively.

We have consolidated the computed values for $m_i^*$, $m_d$, $C_{2D}$, $E_1$, and $\mu$ at 300 K, presenting them

in Table S1. The determination of $E_1$ involves subjecting the system to both compressive and tensile strains, followed by a linear fitting of CBM and VBM values for electrons and holes, respectively (refer to Figure S1(a)). Subsequently, $C_{2D}$ is derived using a quadratic fitting approach applied to the total energy data in response to compressive and tensile strains, as illustrated in Figure S1(b).

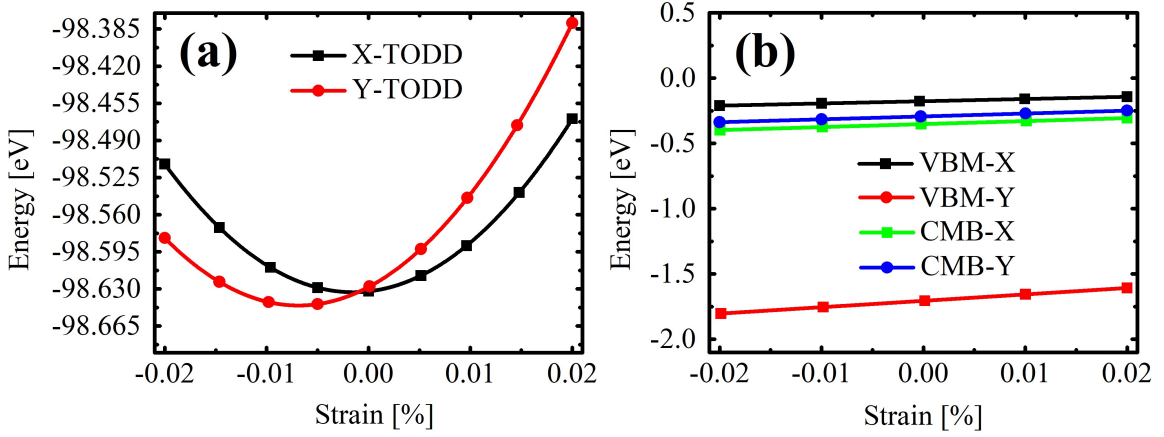

FIG. S1. Panel (a) depicts the quadratic fitting curve for the TODD-G total energy concerning the applied compressive and tensile strains, enabling the derivation of the in-plane stiffness ($C_{2D}$). Meanwhile, Figure (b) presents the linear fitting diagram for the TODD-G total energy concerning the compressive and tensile strains, facilitating the determination of the deformation potential constant ($E_1$).

In TODD-G, the effective masses for holes ($m_h^*$) and electrons ($m_e^*$) are $1.04m_0$ and $0.46m_0$ in the x-direction and $1.13m_0$ and $0.49m_0$ in the y-direction, respectively, with $m_0$ denoting the electron mass. It is worth mentioning that the $m_h^*$ and $m_e^*$ values observed in TODD-G are similar to those reported for pristine and bilayer graphene (about $0.012$-$0.43m_0$ for electrons) [2–4].

TABLE S1. Calculated effective mass $m_i^*$, average effective mass $m_d$, in-plane stiffness $C_{2D}$, DP constant $E_1$, and charge carrier mobility $m_i$ for TODD-G. Here, $m_0$ is the effective mass of a free electron.

| System | Carrier | $m^*$ ($m_0$) | $m_d$ ($m_0$) | $C_{2D}$ (eV/Å) | $E1$ (eV) | $\mu$ ($10^3$ cm$^2$ V$^{-1}$ s$^{-1}$) |
|---|---|---|---|---|---|---|
| TODD | e | 0.46 | 0.37 | 7.67 | 2.29 | 89.25 |
| x-direction | h | 1.04 | 0.47 | 7.67 | 1.70 | 11.13 |
| TODD | e | 0.49 | 0.37 | 7.87 | 2.21 | 77.23 |
| y-direction | h | 1.13 | 0.47 | 7.87 | 4.92 | 10.16 |

The electron mobility outperforms the hole mobility, as highlighted in Table S1. The carrier mobility within the TODD-G monolayer displays anisotropic behavior, showcasing differences between the x and y directions. In the case of TODD-G, the carrier mobility along the x direction exceeds the one along the y direction. This distinction primarily arises from the smaller effective mass of carriers in

the x direction compared to the y direction.

---

[1] J Bardeen and WJPR Shockley. Deformation potentials and mobilities in non-polar crystals. *Physical review*, 80(1):72, 1950.

[2] E Tiras, S Ardali, T Tiras, E Arslan, S Cakmakyapan, O Kazar, Jawad Hassan, Erik JanzÈn, and E Ozbay. Effective mass of electron in monolayer graphene: Electron-phonon interaction. *Journal of Applied Physics*, 113(4), 2013.

[3] K. Zou, X. Hong, and J. Zhu. Effective mass of electrons and holes in bilayer graphene: Electron-hole asymmetry and electron-electron interaction. *Phys. Rev. B*, 84:085408, Aug 2011.

[4] AZ AlZahrani and GP Srivastava. Graphene to graphite: electronic changes within dft calculations. *Brazilian Journal of Physics*, 39:694–698, 2009.



\end{document}